# Coarse Brownian Dynamics for Nematic Liquid Crystals:

# Bifurcation Diagrams via Stochastic Simulation

C. I. Siettos[1], M. D. Graham[2] and I. G. Kevrekidis[1,*]

[1]Department of Chemical Engineering, Princeton University, Princeton, NJ. 08544

[2]Department of Chemical Engineering, University of Wisconsin, Madison, WI

## ABSTRACT

We demonstrate how time-integration of stochastic differential equations (i.e. Brownian dynamics simulations) can be combined with continuum numerical bifurcation analysis techniques to analyze the dynamics of liquid crystalline polymers (LCPs). Sidestepping the necessity of obtaining explicit closures, the approach analyzes the (unavailable in closed form) "coarse" macroscopic equations, estimating the necessary quantities through appropriately initialized, short "bursts" of Brownian dynamics simulation. Through this approach, both stable and unstable branches of the equilibrium bifurcation diagram are obtained for the Doi model of LCPs and their "coarse stability" is estimated. Additional macroscopic computational tasks enabled through this approach, such as coarse projective integration and coarse stabilizing controller design, are also demonstrated.

## INTRODUCTION

Liquid crystalline polymers (LCPs) are large molecules that generally contain long rigid or semirigid segments. Because of these rigid units, they can display phase transitions between isotropic and highly oriented (nematic) states as temperature or concentration is changed. These materials have desirable properties for applications (such as high modulus in the solid phase but low viscosity in the melt) and display a rich variety of phase behavior, especially under flow[1].

Rheological predictions of the behavior of complex fluids like these, often start with the derivation of macroscopic, approximate equations for quantities of interest (order parameters) using various (frequently *ad hoc*) closure approximations; one then brings to bear, on these closed equations, general mathematical techniques for the computation, stability, and parametric analysis of evolution equations (Ordinary or Partial Differential Equations, ODEs or PDEs). The difficulty in obtaining accurate closures has motivated the extensive, in recent years, use of direct simulations, either of the PDE governing the orientation distribution function, or of the equivalent stochastic differential equation, via "Brownian dynamics" (BD) simulations. The latter have the advantage that they are amenable to use with models with many internal degrees of freedom (as opposed to the PDE approach in which the "curse of dimensionality" precludes realistic computation[2]. BD methods have begun to be used in combination with computational fluid dynamics methods to simulate the behavior of complex fluids in spatially inhomogeneous flows[3-14]. In particular, (see Ref. 12) a "lift-run-restrict" procedure like the one described below can be used to simulate spatially inhomogeneous problems for a limited number of ensemble-averaged quantities, thus dramatically reducing the number of PDEs that need to be integrated. While direct temporal simulation tasks can thus be successfully performed, such stochastic simulations are not directly amenable to bifurcation analysis, which is a more appropriate tool for the determination and characterization of the long-term, coarse-grained (we use the term "coarse") macroscopic dynamics and their parameter dependence. Such computations remain, in some sense, limited to continuum macroscopic models.

In this paper we demonstrate a computer-assisted approach that aspires to bridge macroscopic numerical analysis techniques for the (unavailable) closed equations *directly* with microscopic stochastic simulation codes (microscopic/stochastic timesteppers). This system identification based, computational "closure on demand" approach sidesteps the necessity of deriving good *explicit* closures. It enables state-of-the-art microscopic codes, by wrapping a computational superstructure around them, to perform tasks they have not in principle been designed for. Coarse bifurcation analysis, coarse control and coarse projective integration are three such tasks we will illustrate here; the focus will be on the former.

We choose a simple model for the dynamics of LCPs as a prototype with which to illustrate our computational approach. For a quiescent solution of rigid rod molecules, a simple model of the dynamics of the single particle orientational probability distribution function $\psi(\mathbf{u})$ is given by the Smoluchowski equation

$$\frac{\partial \psi(\mathbf{u})}{\partial t} = D\frac{\partial}{\partial \mathbf{u}} \cdot \left( \frac{\partial \psi(\mathbf{u})}{\partial \mathbf{u}} + \psi(\mathbf{u}) \frac{\partial}{\partial \mathbf{u}} \left( \frac{V[\psi, \mathbf{u}]}{kT} \right) \right) \quad (1)$$

where $\mathbf{u}$ is a unit vector describing orientation, $\frac{\partial}{\partial \mathbf{u}}$ is the gradient operator restricted to the unit sphere, $k$ is Boltzmann's constant, $T$ is absolute temperature, $D$ is the rotational diffusivity, here set to unity, and $V[\psi, \mathbf{u}]$ is a nematic potential, a functional of the distribution function that describes the free energy associated with a molecule with orientation $\mathbf{u}$ interacting with its neighbors purely through excluded volume forces[1,15]. We use the simple Maier-Saupe potential

$$V[\psi, \mathbf{u}] = -\frac{3}{2} U \mathbf{uu} : \mathbf{S}, \quad (2)$$

where $\mathbf{S} = \langle \mathbf{uu} \rangle - \frac{1}{3}\mathbf{I}$ is the *tensor* order parameter and $\langle f(\mathbf{u}) \rangle = \int \psi(\mathbf{u}) f(\mathbf{u}) d\mathbf{u}$. The parameter $U$ (the intensity of the nematic potential) can be thought of as proportional to the concentration of the rods. If $\lambda$ is the eigenvalue of $\mathbf{S}$ with the largest magnitude, the so-called *scalar order parameter* $S$ is given by $S = 3\lambda / 2$. The isotropic phase is represented by $S = 0$; it is straightforward to show that this state exists for a quiescent solution for all values of $U$. however when $U > U_{cr} = 5$ this state becomes unstable, and a numerical method is required to analyze the full nonlinear behavior of the equation[15, 16]. It is the resulting bifurcation behavior, the bifurcation diagram of *the stochastically estimated order parameter*, that, as we demonstrate, can be captured by stochastic simulation methods.

The evolution of the distribution function can also be represented by a stochastic (integro-) differential equation[17]:

$$d\mathbf{u} = (\mathbf{I} - \mathbf{uu}) \cdot \left( -\frac{D}{kT} \frac{\partial V}{\partial \mathbf{u}} dt + \sqrt{2D} d\mathbf{w} \right) \tag{3}$$

where $d\mathbf{w}$ is a Wiener process with covariance $\mathbf{I}dt$. It is straightforward to time-integrate this equation with stochastic simulation techniques. We solved it numerically for an ensemble of trajectories $\{\mathbf{u}\} = \mathbf{u}_i(t) : i = 1, N_t\}$ and ensemble averages $\langle f(\mathbf{u}) \rangle$ were evaluated as $\frac{1}{N} \sum_{i=1}^{N_t} f(\mathbf{u}_i)$. We use here an explicit Euler method used in previous studies of liquid crystalline polymers (e.g. Ref. 17).

$$\mathbf{u}_i(t + \Delta t) = \frac{\mathbf{u}_i(t) - \frac{D}{kT} \left. \frac{\partial V}{\partial \mathbf{u}} \right|_t \Delta t + \sqrt{2D\Delta \mathbf{w}_i}}{\left\| \mathbf{u}_i(t) - \frac{D}{kT} \left. \frac{\partial V}{\partial \mathbf{u}} \right|_t \Delta t + \sqrt{2D\Delta \mathbf{w}_i} \right\|} \tag{4}$$

where $\Delta\mathbf{w}$ has zero mean and covariance $\mathbf{I}\Delta t$. To obtain first order weak convergence (i.e. convergence of ensemble-averages) in $\Delta t$, it is not required that $\Delta\mathbf{w}$ be Gaussian[18]; we sample from a uniform distribution.

To perform bifurcation analysis with this stochastic process, we actually analyze the evolution of ensemble averaged properties. The *coarse time-stepper* consists of

(a) The choice of an order parameter at the level of which we believe that a coarse deterministic evolution equation exists and closes. We choose as our order parameter the quantity $\frac{3}{2}(\langle u_z^2 \rangle - \frac{1}{3})$, which, in a slight abuse of notation, we will also call S.

(b) The choice of a (nonunique, one-many) *lifting* operator $\mu$ that maps the macroscopic description S to one or more microscopic descriptions consistent with it. Here the "microscopic" detailed description is the distribution $\mathbf{u}$. The lifting step constructs $\mathbf{u}(S_0)$ distributions *conditioned on* a given $S_0$; our particular choice of lifting is described in more detail below.

(c) The evolution (through the stochastic integrator) of the "lifts" $\mathbf{u}(S_0)$ for a *short macroscopic* time horizon T. This time is associated with the existence of a spectral gap (a separation of time scales in the Smoluchowski equation) and will be further discussed below.

(d) The choice of a *restriction* operator $\boldsymbol{M}$ from the microscopic description (distributions) to the corresponding macroscopic description (moments of the final distributions, possibly averaged over several initial distributions).

The combination of these steps gives us the *coarse* timestepper: an estimation of $S(t{=}T)$, the result of integrating the (unavailable in closed form) equation for S with initial condition $S\ (t{=}0)$ for time T; the coarse time-stepper is closely related to the optimal predictors of Chorin and coworkers[19]. For the procedure to be practically successful, it is important that a separation of time scales exist in the evolution of the distribution $\mathbf{u}$. In particular, consider a discretization of $\mathbf{u}$ in terms of (a sufficiently large number of) its moments. We expect that, for the conditions of interest, this is a singularly perturbed problem: the discretized system of coupled nonlinear ODEs for the moments evolves quickly to a one-dimensional slow manifold parametrized by S. This slow manifold (which can be thought of as a center manifold, or possibly even an inertial manifold) is a graph of a function over S; all moments quickly become "slaved to" –evolve to become functionals of- S. While the conditions for such a "fast" slaving to occur may not be easy to explicitly write down, or verify in a particular simulation, it is still interesting to present the following heuristic argument. If we work in a regime in which we believe that a deterministic *closed* evolution equation can be written for S *only*, then such a separation of time scales *must be valid*! For, if the higher moments did not quickly evolve to functionals of S, the scalar initial value $S(t{=}0)$ would not be sufficient to deterministically predict S later on in time in a simulation or an experiment: the actual initial values of the higher moments would significantly affect $S(t{=}T)$.

Consider, as an illustration, an isothermal molecular simulation of a practically Newtonian fluid: if the stresses at the initial configuration are not proportional to velocity gradients, they would very quickly become so. Newton's law of viscosity then implicitly defines the "slow manifold", on which fields of higher moments of the molecular distribution are slaved to the lowest two "determining" moment fields: density and momentum. The Navier-Stokes equation then becomes a sort of "Approximate Inertial Form" for the hierarchy of moments of the Boltzmann equation. In a similar sense, in the Smoluchowski equation above is a sort of ``approximate inertial form" for the detailed system Fokker-Planck coarse grained in terms of the single particle orientational probability density.

The coarse variable we "evolve" in our simulation is $\frac{3}{2}(\langle u_z^2 \rangle - \frac{1}{3})\ (\equiv \mathrm{S})$, a simple measure of the degree of orientation. We initialize the orientation vectors as 3-vectors in full space, but constrained to lie on the unit sphere; the *z* (vertical) direction thus determines the north and south poles of the sphere. The most important issue in our computations is the *lifting* step: how to "reconstruct" or "initialize" a full distribution function $\{\mathbf{u}\}$ given a specified value $E$, for $\langle u_z^2 \rangle$. There is clearly no unique solution to this problem (the lifting operator is not unique). A simple strategy that we find to be effective both here and in a related application to spatially varying systems[12] is a minimization with respect to an *a priori* chosen reference ensemble $\{\mathbf{u}_{\mathrm{ref}}\}$. That is, we seek an ensemble $\{\mathbf{u}\} = \{\mathbf{u}_{\mathrm{ref}} + \Delta\mathbf{u}\}$ and determine $\{\Delta\mathbf{u}\}$ by solving the minimization problem:

$$\min \langle \Delta\mathbf{u} \cdot \Delta\mathbf{u} \rangle, \mathrm{s.t.} \left\langle \left(\mathbf{u}_{\mathrm{ref}} + \Delta\mathbf{u}\right)_z^2 \right\rangle = E \qquad (5)$$

This problem can be reduced to a linear least squares problem if the $\Delta\mathbf{u}$ are sufficiently small. For an isotropic reference distribution, it is straightforward to show that the minimal corrections in the linearized problem have no component in the *x* or *y* directions, and the *z* component can be found analytically. Our procedure, then, is to solve the linearized version of (5) for $\Delta u_Z$. Then the *x* and *y* components of $\Delta\mathbf{u}$ are obtained by requiring it to be a unit vector with the same azimuthal angle as the corresponding reference vector (in spherical coordinates where the polar angle is measured from the *z* axis). We note that this "lifting" procedure always yields distributions that are (statistically) axisymmetric with respect to the *z* axis; such distributions constitute an invariant, though not necessarily stable, subspace for the Smoluchowski problem. As we will see below, we have also found it useful to construct initial ensembles conditioned not only on the value of $\left\langle u_z^2 \right\rangle$, but also on the *standard deviation* $\sigma = \left( \frac{1}{N-1} \sum_{i=1}^{N} (u_{zi} - \overline{u_{zi}})^2 \right)^{1/2}$. After initializing a distribution conditioned on S, additional particles were inserted to impose the desired variance while preserving the value of S.

A final note before moving on to the results. This problem is in fact a highly degenerate one – there are two continuous (rotational) symmetries and no preferred direction, so it is O(3) equivariant[20]. The trivial (isotropic random orientation of molecules) solution is spherically symmetric – invariant under polar or azimuthal rotations. The states that bifurcate from this can have arbitrary orientation, so S is only unique to within a pair of rotations. Bifurcation problems of this nature are actually rather difficult to numerically study, unless the algorithm "knows" about the symmetries. Here, the restriction operator constrains the solution behavior to a one-dimensional subspace of the global stable manifold, reducing the bifurcation problem to a generic one that is easily treated by standard methods (the invariances have been factored out). Furthermore, the lift operator, because it uses the istrotropic distribution as a reference, keeps initial conditions close to the axisymmetric invariant subspace. The downside to this restriction is that the stability predictions made here are only valid in the subspace considered. Nevertheless, stability in the unrestricted space can be determined straightforwardly upon location of the steady state, by linearization and eigenanalysis in the full space. It is worth mentioning the recent development of a template-based method that allows the dynamic factoring out of symmetry; this method is applicable to general systems with symmetry and even to self-similar dynamical systems[21, 22].

## SIMULATION RESULTS

**Coarse Bifurcation analysis for the nematic Brownian dynamics model.** We found the steady state bifurcation diagram for S as a sequence of fixed points of the coarse timestepper $\Phi_T$:

$$S(t=T) \equiv \Phi_T(S(t=0)) = \boldsymbol{M}\mathbf{U}_T(\mathbf{u}_0) = \boldsymbol{M}\mathbf{U}_T(\mu E)$$

the result of lifting consistently $S(t=0)=E$ to an initial distribution $\mathbf{u}_0 = \mu E$, evolving the stochastic differential equations for the distribution $\{\mathbf{u_0}\}$ over a time interval or reporting horizon T to the final distribution $\mathbf{U}_T$, and restricting back to $S=\boldsymbol{M}\mathbf{U}$. The (coarse) derivatives for the Newton-Raphson fixed point iteration were estimated numerically through application of the coarse timestepper to nearby S initial conditions. For higher dimensional systems, Newton-Krylov type methods are used, based on iterative identification of the slow subspace of the linearization of the timestepper[23]. Stability (the leading part of the spectrum of the linearization of the dynamic problem at stationarity) can be deduced from the linearization of the coarse timestepper at its fixed point. Here this is a scalar quantity -- a single "multiplier" for the discrete-time system, from which a single "exponent" (the eigenvalue of the corresponding continuous time system) is deduced. In problems with more degrees of freedom, it is the leading (slow) part of the Jacobian that is approximated as a byproduct of procedures like the Recursive Projection Method of Shroff and Keller[24]. If the steady state is known, then matrix free algorithms like an Arnoldi procedure can be used to estimate the slow subspace of the coarse timestepper. This was first used in macroscopic flow computations by Christodoulou and Scriven[23] and in viscoelastic computations by Ramanan and Graham[25]. Somasi and Khomami have used transient microscopic simulations to quantify coarse stability[10,11].

The coarse timestepper based fixed point algorithm (essentially a Newton-Raphson) was combined with arclength continuation and branch-switching algorithms. The coarse bifurcation diagram of the order parameter, the largest eigenvalue of the second-rank order tensor $\mathbf{S}$ w.r.t the potential intensity $U$ is shown in Figure 1. It was computed (upon convergence of the Newton-Raphson to a residual of $O(10^{-4})$, for perturbations $\varepsilon \sim 10^{-2}$), using the microscopic Brownian dynamics simulator. The computational parameters were: number of trajectories $N_{traj} = 3\text{x}10^5$, time-reporting horizon T=1.75 and an inner Euler integrator step set to $\Delta t=0.0005$ (the sensitivity of the results to the latter was also carefully monitored). These results should be compared with Figure 1 of Faraoni *et al*[16], where the bifurcation diagram is computed from a standard bifurcation analysis of a spherical harmonics Galerkin expansion of the Smoluchowski equation.

The isotropic ("flat") solution loses stability at what appears (for our scalar coarse variable) like a transcritical bifurcation at a critical potential $U_{cr}$, giving rise to two partially oriented anisotropic solutions. The predicted critical value of $U_{cr}$ as calculated with our "Coarse Brownian Dynamics" procedure was found to be $U_{cr} = 5.01$ and agrees within 0.2% with the predictions of the Smoluchowski equation. A turning point was found to be on the subcritical prolate ($S > 0$, nematic) branch at $U^* \approx 4.6$ (within 2.2% of the predictions obtained using the discretized Smoluchowski equation). The stability of the linearized system is monitored by computing the norm of the eigenvalues that cross the unit circle. The stability results here are consistent with those dictated by bifurcation theory: solutions on the subcritical prolate branch are unstable between $U^*<U< U_{cr}$ and regain stability past the turning point (for $U>U^*$); solutions on the oblate branch ($S < 0$) *appear* stable in these computations. Extensive time evolution of the stochastic system shows that the oblate branch is indeed unstable with respect to perturbations that drive it to a prolate branch that does not have z-axisymmetry imposed by our lifting step.

It is interesting to consider the *apparent stability* of the computed oblate branch. We know from fully discretized Smoluchowski simulations that this branch is *unstable*. It is actually a saddle branch: most directions in phase space are attracted to the steady states, and only the ones destabilized at $U_{cr}$ are unstable. Close to $U_{cr}$ these unstable modes are so slow, that the BD simulator, depending on its time horizon, does not initially "see" the instability (see Fig.2b,c,d) , considering this direction as practically neutral. It records the movement along one of the slowest attracting direction, and reports the steady state as stable. Of course, if we let the time-reporting horizon of the time-stepper grow longer, the instability will be correctly characterized. This might appear at first sight as a defect of the approach; on the contrary, we believe that it may be a strong point. What the coarse timestepper reports, is the expected behavior *over the simulation ensemble and time-horizon chosen*. It is well known that different apparent dynamics occur when one studies a noisy phenomenon at different time levels. Instead of deriving different equations for the expected behavior over different time scales, we can in principle analyze the behavior of several such "layers" using the *same* inner detailed simulator, but tailoring the simulation ensemble (initial conditions and observation time) to the "layer" of interest (see Ref. 26 for a discussion).

Figure 3 illustrates the computed exchange of stability at $U_{cr}$; triangles correspond to the eigenvalues calculated on the flat branch while rhombs correspond to the ones calculated on the branch containing the turning point. (The above coarse bifurcation analysis is based on the hypothesis that a macroscopic coarse model exists *and closes* for S, a single statistic of the underlying microscopic distribution. This implies that higher-order moments of the **u** distribution become quickly *slaved* to lower ones (they evolve towards a "slow manifold" parametrized by the lower ones). Computational results corroborating this can be seen in Fig 4**,** which illustrates (both in terms of the direct and of the cumulative distribution) the "initial fast" and "subsequent slow" evolution stages.

A "phase portrait" of the trajectories of a few different evolving distributions is seen in Fig. 5, where two moments of the distributions are plotted: S as well as $\sigma$, the standard deviation of the microscopic distribution in the z direction (which is a "higher" than S moment in the corresponding hierarchy). A one-dimensional slow manifold, parametrized by S is clear in the picture; all three coarse steady states clearly lie on this manifold.

It is interesting to consider the *transient approach* of various initial distributions to this manifold. Three time scales exist at this resolution: (a) an initial very fast collapse onto a two dimensional slow manifold parametrized by S and $\sigma$; this is the "healing" of the errors made in the lifting step[26-29]; (b) this is followed by a somewhat slower approach to the one-dimensional slow manifold, parametrized by S; and finally (c) a "slow" approach to the ultimate steady state on this slow manifold.

It is interesting that when we "lift" (i.e condition the microscopic distributions on two coarse variables, S and $\sigma$) we have a two-dimensional timestepper, and its fixed points, upon convergence of the Newton-Raphson to a residual of $O(10^{-4})$ for $\varepsilon \sim 10^{-2}$, possess two eigenvalues (multipliers), representative of the corresponding relaxation times. In the particular case shown (at $U$=4.75 using $N_{traj} = 10^5$) the two eigenvalues were found to be $\lambda_1$=0.18 (the corresponding eigenvector is depicted with the red dotted line) and $\lambda_2$=0.001 (the corresponding eigenvector is depicted with the blue dotted line). This implies that the (discrete time) evolution of the second "coarse mode" is about 500 faster than the first "coarse mode", the slowest, governing one. Notice that the eigenvector corresponding to the slow eigenvalue is aligned with the (visually apparent) one dimensional slow manifold. Our computational results at this level of accuracy were converged with respect to the number of trajectories and the inner step of the Euler microscopic integrator.

It makes sense to begin such computations in a regime where we know at least at what level one might obtain closures (at these values of $U$ and in the absence of shear, it was known that one can close with S only). What happens as we approach (in a continuation environment) conditions where the closure will fail? The situation is discussed in detail in Ref. 26: moments "higher up" in the hierarchy, that were fast enough for lower $U$, start becoming slow. We must then augment the set of independent coarse variables; the same "lift-run-restrict" procedure can be used for coarse computation as long as we simply "lift" (construct distributions conditioned on) *a higher number of moments* as independent coarse variables (e.g. see Ref.12). Performing such a check regularly along a continuation branch (keeping track of whether the "next fastest" mode is still fast enough to get slaved over our reporting horizon) is the analog, in our case, of checking from time to time whether the mesh for a given discretization problem needs refinement or not.

**Coarse Control for the nematic Brownian Dynamics model.** The proposed computational framework serves as a "just in time" or "on demand"[30] computational closure methodology that allows the identification, from short computational experiments, of coarse time derivatives, "coarse slow" Jacobians, coarse derivatives with respect to parameters etc. These quantities are used in conjuction with traditional, continuum scientific computation to find coarse fixed points. These fixed points, along with coarse linearization around them are precisely the "systems level" information by a linear control design algorithm. It becomes then possible to invoke such algorithms and design coarse observers and controllers that will stabilize coarse unstable stationary states; extensions to nonlinear control are straightforward.

For demonstration we designed a stabilizing controller for the macroscopic *unstable* stationary state at $U_0 = 4.7$. This coarse steady state (evaluated through the T=1.75 coarse Brownian timestepper, upon convergence of the Newton iteration to a residual of $O(10^{-5})$ for $\varepsilon \sim 10^{-2}$) is $S_0 \approx 0.15$. It is assumed that the discrete model describing the system behavior around the equilibrium is given by the standard discrete time stochastic state-space model: $\mathbf{x}(k+1) = \mathbf{A}\mathbf{x}(k) + \mathbf{B}\mathbf{u}(k) + \mathbf{w}(k)$; $\mathbf{w}(k)$ denotes the process noise.

At the steady state, the estimates of the (scalar, since the problem is one-dimensional) Jacobian $\mathbf{A}$ and control matrix $\mathbf{B}$ were found to be $\mathbf{A} \approx 1.53$ and $\mathbf{B} \approx 0.33$. Actually this information is a byproduct of the fixed point/continuation procedure for estimating the location of the coarse steady states. At this point we should note that for coarse large-scale problems (such as those arising in discretized coarse PDEs), Newton-Picard-type algorithms (e.g the RPM algorithm) can be used to derive the "coarse *slow*" Jacobian matrices. We employ our bifurcation parameter $U$, the intensity of the nematic potential, as the control actuator. We used a linear feedback controller of the form $U(t) - U_0 \equiv u(t) = -K(S - S_0)$. For our illustrations, we aimed at stabilizing the unstable coarse steady state, by placing the coarse eigenvalue to $\lambda \approx 0.95$ with a sampling time DT=dt=0.005; the required control gain was found to be $K \approx 54$. The number of trajectories was set to $N_{traj} = 3\text{x}10^3$. If necessary, a Kalman filter could also have been constructed based on the coarse information[31]. The open and closed loop responses are shown in Fig. 6.

**Coarse Projective Integration of the Brownian Dynamics model.** We have demonstrated that our coarse "lift-run-restrict" procedure can be used to enable the performance of several numerical tasks (bifurcation, continuation, stability analysis and control) directly at the macroscopic level, sidestepping the necessity to obtain explicit macroscopic closures. We will now briefly demonstrate that this procedure can also be used to accelerate the time evolution computations directly. The method to accomplish this, called "Coarse Projective Integration" is described in more detail in References 29 and 32. Here we only demonstrate its simplest realization (the coarse forward Euler projective method).

The main idea is to consider the computations that would have been performed, had an explicit closure (in the form of a scalar ODE in the form $\frac{dS}{dt} = f(S)$. Given an initial condition $S(t=0) = S_0$ a simple, explicit, forward Euler integration (with time step $\Delta t$) would give $S(t=\Delta t) = S_0 + \Delta t f(S_0)$. Since the time derivative functional form $f(.)$ is not explicitly available, we *estimate* it from short duration BD simulations initialized consistently with $S_0$. The steps of the algorithm are then as follows (given the lifting and restriction operator choices we discussed above):

(a) Select an initial condition $S_0$

(b) Lift it to one (or more) consistent microscopic distributions, $\mathbf{u}_0 = \mu S_0$

(c) Evolve microscopically for enough time $\mathbf{t}_1$ for the lifting errors to heal, and restrict to $S_1 = \boldsymbol{M}(\mathbf{u}(t=t_1))$; evolve a little longer, till time $t_2$, restrict to $S_2 = \boldsymbol{M}(\mathbf{u}(t=t_2))$.

(d) Use the difference $\frac{S_2 - S_1}{t_2 - t_1}$ to estimate the derivative $\frac{dS}{dt}\Big|_{t=t_2}$

(e) Project in the future to an estimate $S(t = t_3) = S_2 + (t_3 - t_2)\frac{S_2 - S_1}{t_2 - t_1}$.

(f) Return to step (b)

One can clearly see an "inner" integrator (the BD simulator) and an "outer" integrator (a Forward Euler method, that uses the results of the inner integrator). This is the simplest form of the Coarse Projective Forward Euler method, with simple differencing for the estimation of derivatives; much more sophisticated components can be used in assembling such multilevel integration schemes, but the idea remains the same. The numerical analysis of these algorithms is the subject of extensive research[36], as are the issues of modifying them for the case where the "inner" evolution code is a microscopic/ stochastic (here, the BD simulator). Our purpose here is not to analyze these algorithms, but just to illustrate their underlying principle. This is accomplished in Fig. 7, where short bursts of BD simulation (marked with red asterisks) are used to estimate (after an initial transient) the time derivative $\frac{dS}{dt}$. This quantity is then used to perform a "projection" into the future (predict the expected value of $S$ after some time. The procedure is then repeated: we lift from the predicted value of S to a new microscopic distribution, run for some time to obtain a new estimate of $\frac{dS}{dt}$, project again and so on. For comparison, we have included in the Figure 7 the projection (on S) of a long, uninterrupted BD transient. It is clear that, in this case, the procedure eventually saves us 60% of the BD simulation flops (the projection time interval is 3/2 the BD evolution one) ! The errors made when lifting from the "projected in time" value of S are seen to quickly "heal" as the BD simulation is restarted. This is a consequence of the exponential attractivity of the "slow manifold" parametrized by our coarse variable S.

**Discussion and Conclusions**. We presented and illustrated a computational methodology for the coarse, multiscale computational study of microscopic stochastic simulators. Our example was the “enabling” of Brownian Dynamics simulators of nematic liquid crystal models to perform macroscopic tasks such as the location of stable and unstable coarse stationary states, their stability, continuation and bifurcation analysis, as well as additional tasks (controller design, coarse projective integration etc.). We believe that these computer-assisted techniques, grounded in the power of an “inner” microscopic simulator, and based on the *conceptual* existence of a macroscopic closure, offer the promise of a new bridge across the scale gap, between “the best available” microscopic/stochastic simulators and their macroscopic, coarse dynamics. Based on the separation of time scales that fundamentally underlies macroscopic deterministic equations, these algorithms sidestep the derivation of explicit equations, but do allow the use of a large arsenal of “equation-based tools”, developed for continuum models, to be used directly on the microscopic solvers. Many “systems” tools, ranging from system identification and filtering, to variance reduction and matrix free iterative linear algebra methods form part of this bridge. An extensive discussion of the overall approach can be found in Ref. 29.

What we presented here was a simple *illustrative* example: we applied the method on a stochastic realization of a *known* Smoluchowski equation, so as to demonstrate its salient features in a case where the “correct dynamics” were known. It is important, however, to point immediately out that the method can be used in precisely the same way when the Smoluchowski equation is *not accurately known* (e.g. in the case of molecules comprised of several bead-spring segments) or in cases where the inner solver may be of a different nature (e.g. MD as opposed to BD).

The fundamental underlying principle of this “computational enabling technology” is the smoothness, or regularity of the expected behavior with respect to time (allowing us to estimate coarse temporal derivatives in projective integration) and with respect to the variables themselves (so that the action of coarse slow Jacobians can be estimated for bifurcation computations). It is worth mentioning that a similar regularity of the coarse behavior *but now in macroscopic space* allows, under certain circumstances, the lifting to be performed not over a full computational domain, but only over small computational “patches”. The microscopic simulation is then performed over such patches (communicating with each other through a coarse macroscopic field only[12,29]. The conditions under which one can exploit such “short space”-“short time” microscopic simulations, using regularity in space and time to interpolate coarse macroscopic fields and projectively integrate them in time, is discussed in detail in Ref.29.

**Acknowledgements** This work was partially supported by NSF (through the ITR and the Nanoscale Modeling and Simulation Programs) and the AFOSR. It is a pleasure to acknowledge a long-standing collaboration with Prof. C. W. Gear, with whom coarse projective integrators were jointly developed. We are indebted to Profs. G. Forest, R.C. Armstrong, H.-C. Oettinger, and to Drs. Gopinath and Zhou for many helpful discussions.

**REFERENCES**

[1]R. G. Larson, The Structure and Rheology of Complex Fluids (Oxford, New York, 1999).

[2]N. Gershenfeld, The Nature of Mathematical Modeling (Cambridge, Cambridge, 1999).

[3]M. Laso and H. C. Öttinger, J. Non-Newtonian Fluid Mech. ,**47**, 1 (1993).

[4] M. A. Hulsen, A. P. G. van Heel and B. H. A. A. van den Brule, J. Non-Newtonian Fluid Mech., **70**, (1997); also A. P. G. van Heel and M. A. Hulsen and B. H. A. A. van den Brule, J. Rheol., **43**, 1239 (1999).

[5]T. W. Bell, G. H. Nyland, J. J. de Pablo and M. D. Graham, Macromolecules, **30**, 1806 (1997).

[6]P. Halin, G. Lielens, R. Keunings and V. Legat, J. Non-Newtonian Fluid Mech., **79**, 387 (1998).

[7]X. Gallez, P. Halin, G. Lielens, R. Keunings and V. Legat, Comp. Meth. Appl. Mech. Eng., **180**, 345 (1999).

[8]X. J. Fan, N. Phan-Thien and R. Zheng, J. Non-Newtonian Fluid Mech., **90**, 47 (2000).

[9]P. Wapperom, R. Keunings and V. Legat, J. Non-Newtonian Fluid Mech., **91**, 273 (2000).

[10]M. Somasi and B. Khomami, J. Non-Newtonian Fluid Mech., **93**, 339 (2000).

[11]M. Somasi and B. Khomami, Phys Fluids, **13**, 1811 (2001).

[12]R. M. Jendrejack, J. J. de Pablo and M. D. Graham, J. Non-Newtonian Fluid Mech., in press (2002)

[13]C. C. Hua and J. D. Schieber, J. Rheol., **42** , 477 (1998).

[14]J. Cormenzana, A. Ledda , M. Laso and B. Debbaut, J. Rheol., **45**, 237 (2001).

[15]M. Doi, J. Polym. Sci., Polym. Phys. Ed., **19**, 229 (1981).

[16]V. Faraoni , M. Grosso, S. Crescitelli and P. L. Maffettone, J. Rheol., **43**, 829 (1999).

[17]R. G. Larson and H.C. Öttinger, Macromolecules, **24**, 6270 (1991).

[18]P. E. Kloeden and E. Platen,Numerical Solution of Stochastic Differential Equations (Springer, NY, 1992).

[19]A. Chorin, O. Hald and R. Kupferman, Proc. Natl. Acad. Sci. USA, **97**, 2968 (2000)

[20]M. Golubitsky, D. G. Schaeffer and I. Stewart, Singularities and Groups in Bifurcation Theory (Springer, New York, 1985).

[21]C. W. Rowley, I. G. Kevrekidis, J. E. Marsden and K. Lust, Nonlinearity, submitted (2002).

[22]C. W. Rowley and J. E. Marsden, Physica D, **142**, 1 (2000).

[23]K. N. Christodoulou and L. E. Scriven, Quart. Appl. Math. **9**, 17 (1998).

[24]G. M. Shroff and H. B. Keller, SIAM J. Numer. Anal., **30**, 1099 (1993).

[25]V. V. Ramanan, K. A. Kumar and M. D. Graham, J. Fluid Mech., **379**, 255 (1999).

[26]A. G. Makeev, D. Maroudas, and I. G. Kevrekidis, , J. Chem. Phys, **116,** 10083 (2002).

[27]C. Theodoropoulos, Y. H. Qian, and I. G. Kevrekidis, PNAS, **97**, 9840 (2000).

[28]A. G. Makeev, D. Maroudas, Z. Panagiotopoulos and I. G. Kevrekidis, J. Chem. Phys, **117** (2002)

[29]I. G. Kevrekidis, C. W. Gear, J. M. Hyman, P. G. Kevrekidis, O. Runborg and C. Theodoropoulos, submitted to Communications in the Mathematical Sciences (2002). Also can be found as http://www.arxiv.org/PS_cache/physics/pdf/0209/0209043.pdf

[30] Cybenko, G., in *Identification, Adaptation, Learning*, S. Bittani and G Picci Eds pp.423-434 NATO ASI, Springer (1996).

[31]C. I. Siettos, A. Armaou, A. G. Makeev and I. G. Kevrekidis, submitted to *AIChE J.* (2002) Also can be found as http://www.arxiv.org/ftp/nlin/papers/0207/0207017.pdf

[32]C. W. Gear and I. G. Kevrekidis, SIAM J. Sci. Comp. (in press); also NEC Technical Report NECI TR 2001-029, can be found at http://www.neci.nj.nec.com/homepages/cwg/projective.pdf

**Figure Captions**

**Figure 1**. Coarse bifurcation diagram for the nematic model for $N_{traj}=3x10^5$, dt=0.0005, T=1.75; blue rhombs correspond to stable steady states, red ones correspond to unstable steady states. These are obtained as fixed points of the BD timestepper.

**Figure 2:** Evolution of $\left\langle u_z^2 \right\rangle$ for different initial values, intensity potentials and stochastic integrator time steps. (a), (b) $U$ =6.5, $N_{traj}=10^3$, dt=0.001, (c) $U$ =5.5, $N_{traj}=10^3$, dt=0.001, (d) $U$ =5.5, $N_{traj}=3x10^5$, dt=0.0005. The purpose of the figure is to illustrate some of the parametric and numerical effects, but also to show that initial transients may linger enough around an unstable steady state for the coarse timestepper to find it as a fixed point.

**Figure 3**. Exchange of stability at $U_{cr}$; triangles correspond to the eigenvalues calculated on the isotropic branch while rhombs correspond to the ones calculated on the prolate branch.

**Figure 4.** (a) Evolution of the distribution function of **u** (histogram) in the z – direction. The values in the z- direction were partitioned in 100 bins. (b) Evolution of the corresponding cumulative distribution function of **u** in the z – direction. The simulations were performed at $U$= 5.5, $N_{traj}=10^3$, dt=0.001

**Figure 5**. Phase portrait (S, standard variation $\sigma$) at $U$=4.75. The results are obtained through the BD timestepper using $N_{traj}$ =$10^5$ and dt=0.0005; the blue dotted line depicts the eigenvector corresponding to $\lambda_1$=0.001, while the red dotted one depicts the eigenvector corresponding to $\lambda_2$=0.18.

**Figure 6**. Stabilizing an unstable steady state (at $U_0$ =4.7); $N_{traj}$ =$3x10^3$, (a) Open loop responses (green solid lines); Closed loop response (blue solid line), (b) closed response of the control variable $U$; (red dotted lines correspond to nominal values).

**Figure 7**. Coarse integration for the nematic model: Circles correspond direct simulation at $U = 4.7$, $N_{traj}=10^5$, dt=0.0005; required computational time (from time=0 to time=25) on Pentium III, 1.2 GHz: 100 min. Asterisks correspond to the coarse integration procedure: lifting, evolving, recording the restrictions, and then using the estimated time derivative to project (here simply with projective forward Euler) in time. After an initial transient, the projection interval is 1.5 times the detailed BD evolution interval. Total required computational time: 72 min.

.

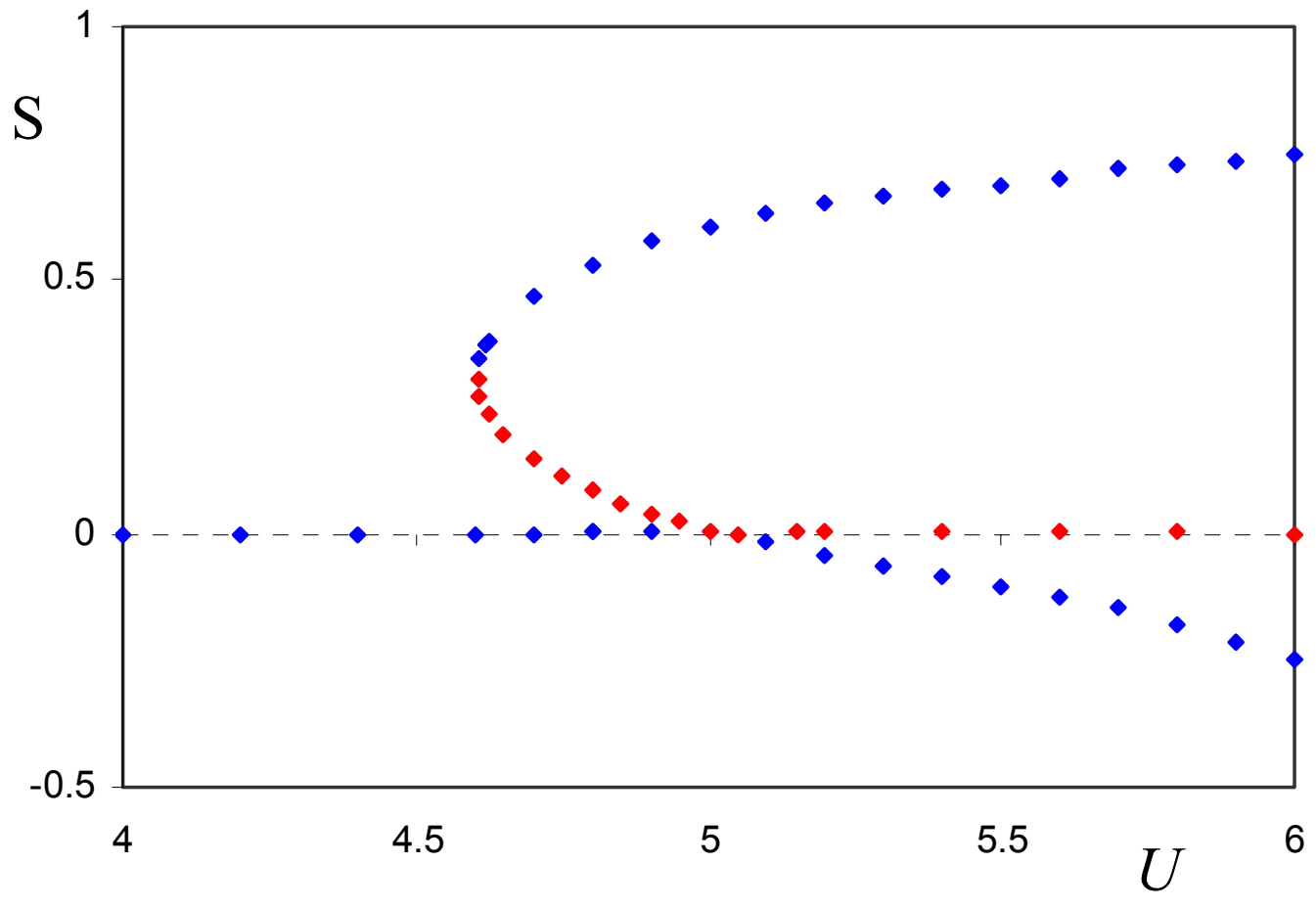

Figure 1.

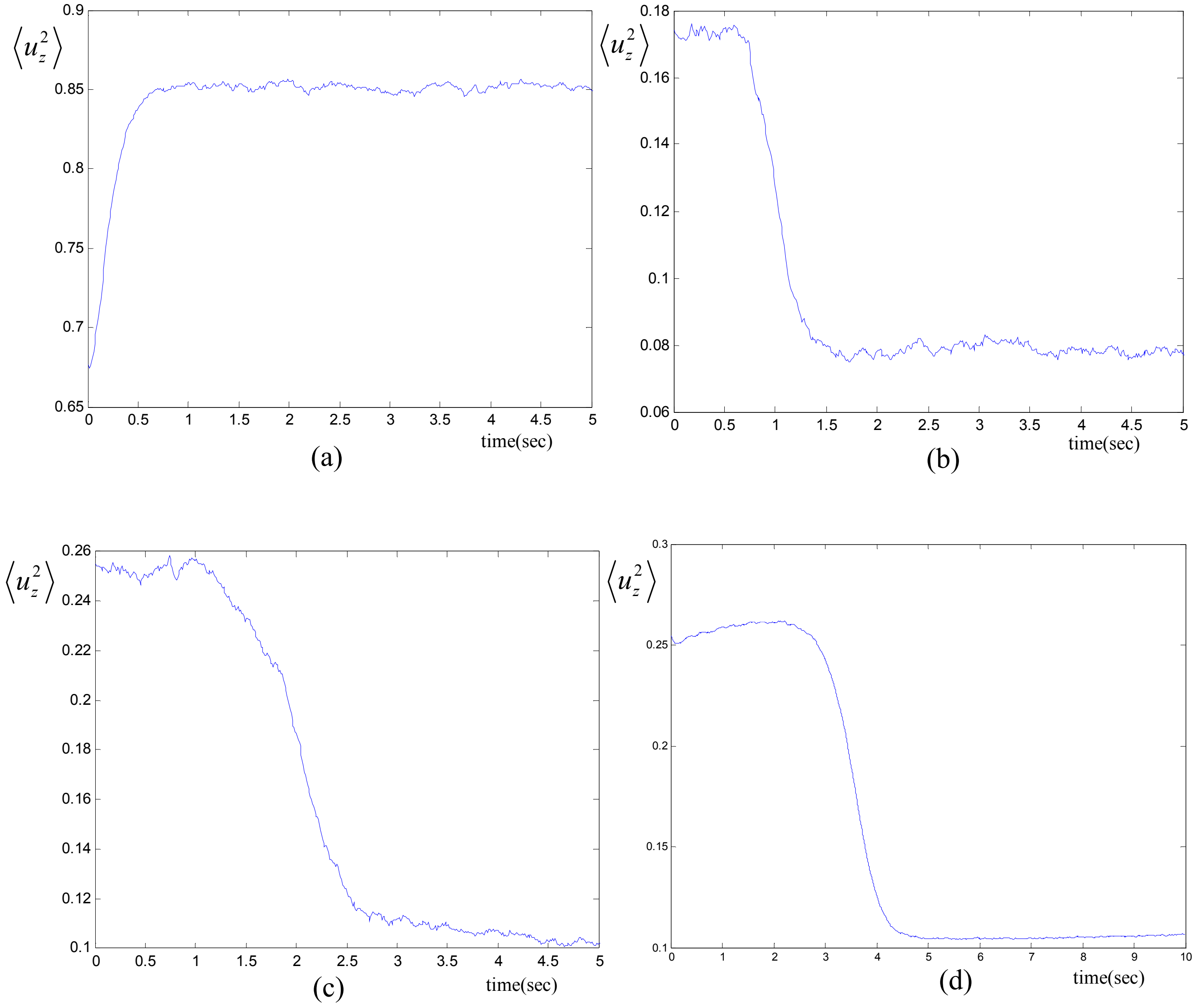

Figure 2.

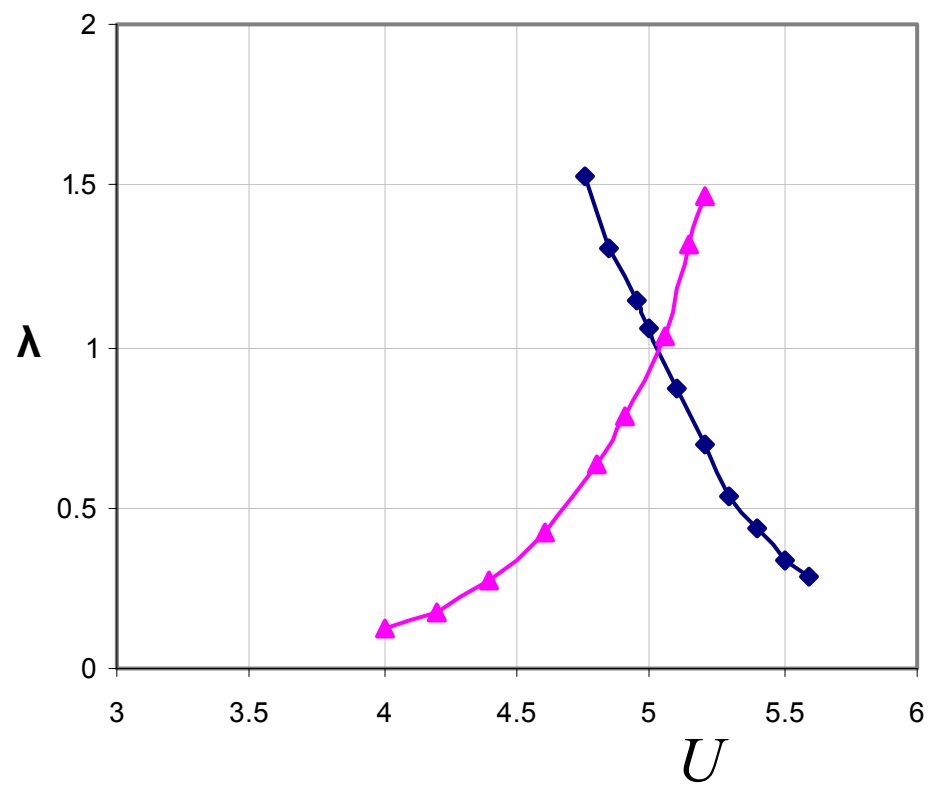

Figure 3.

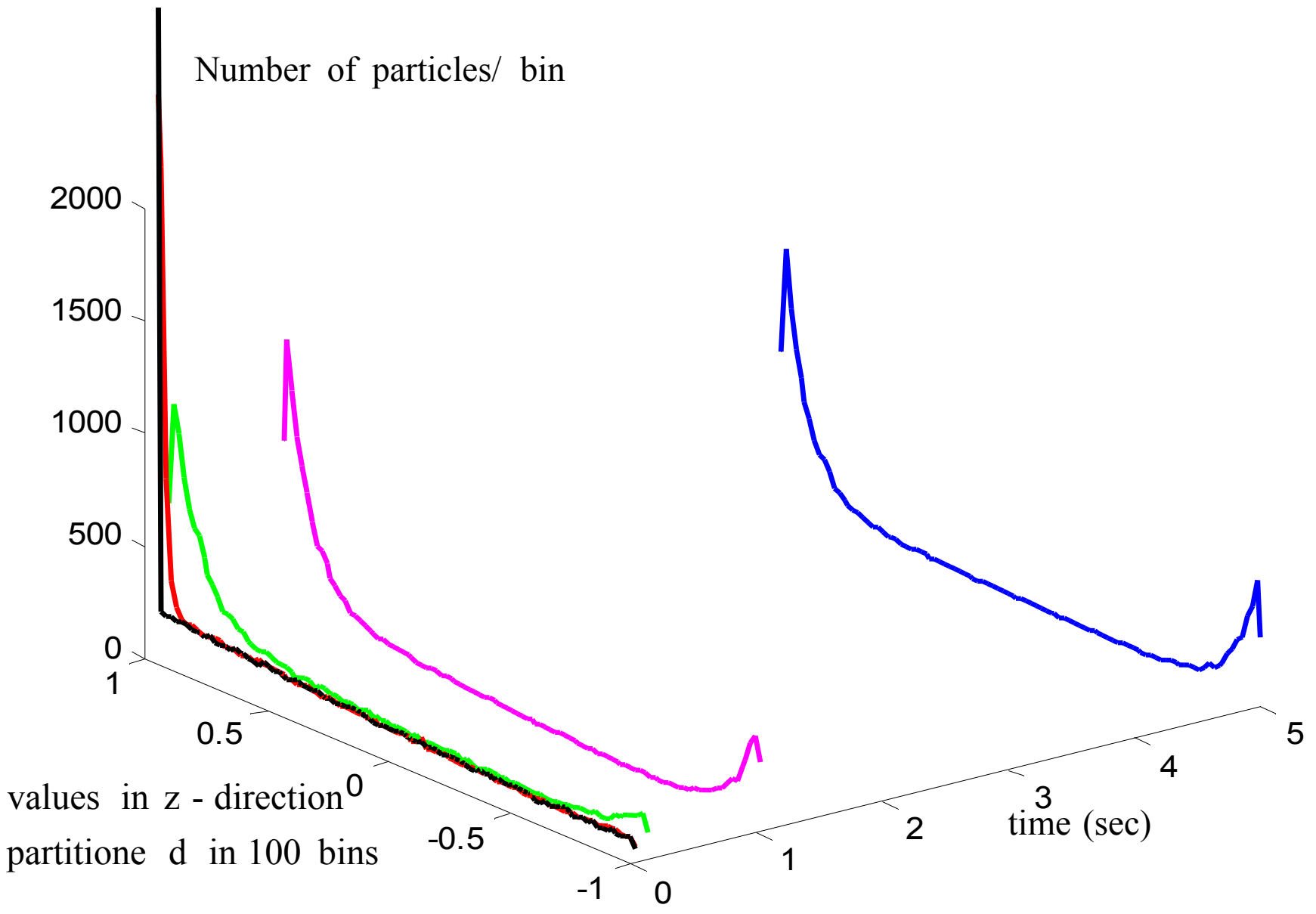

(a)

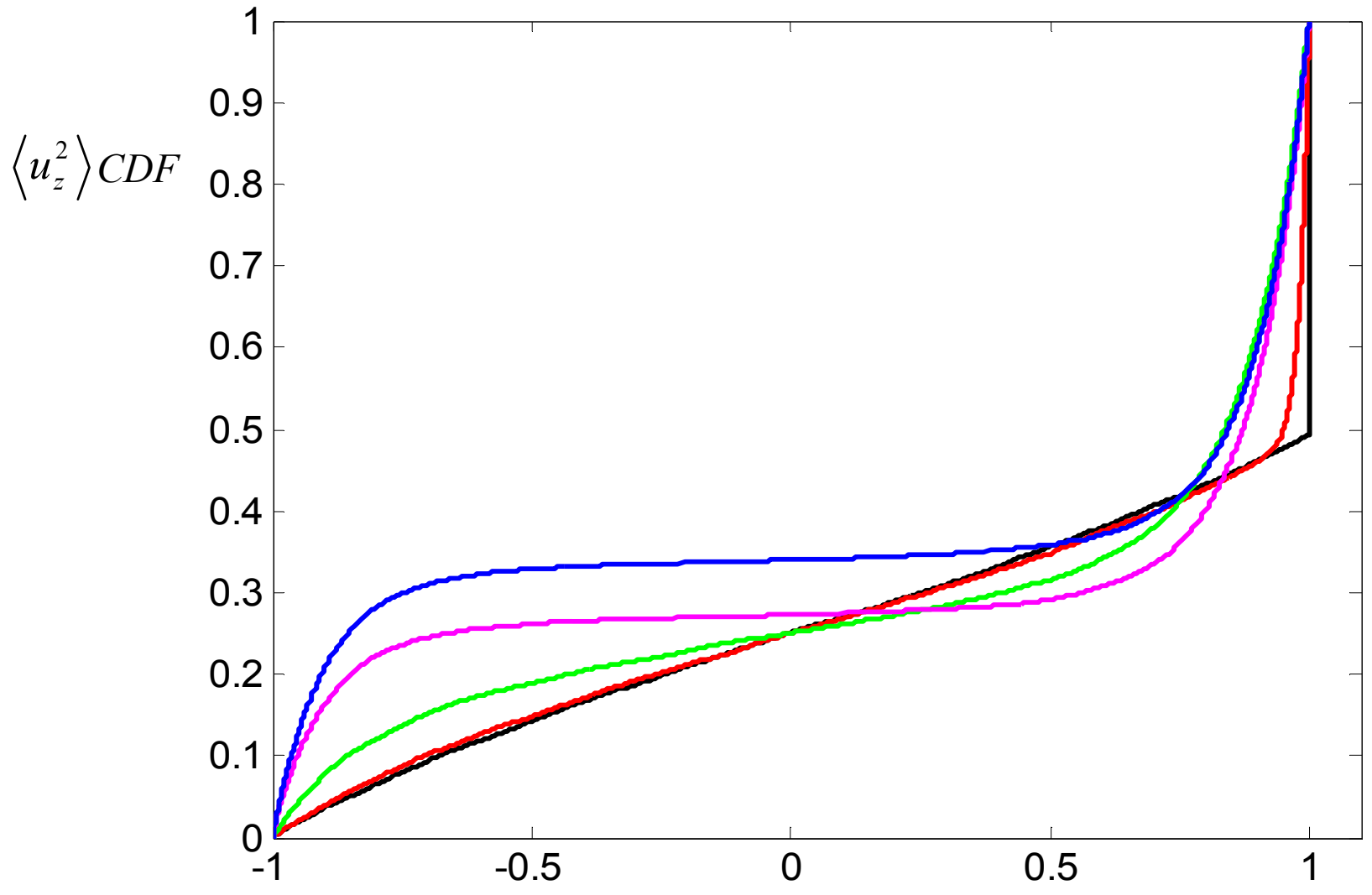

(b)

Figure 4.

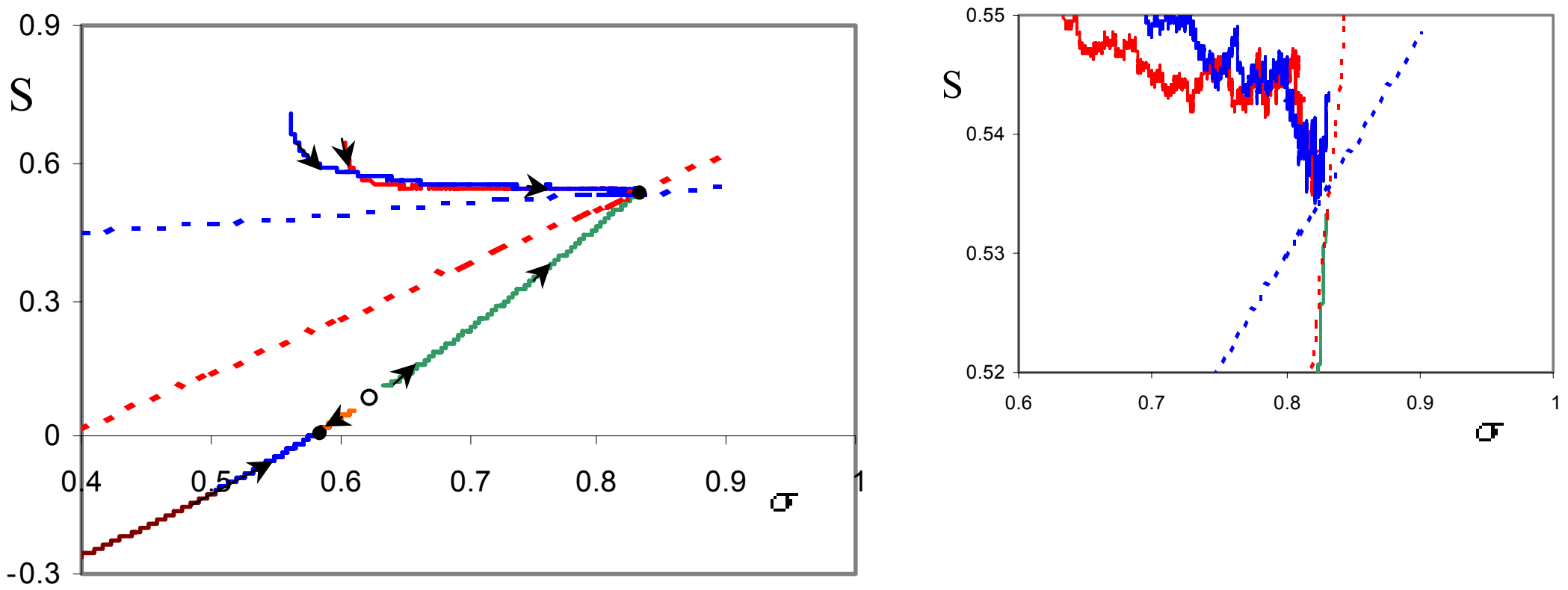

Figure 5.

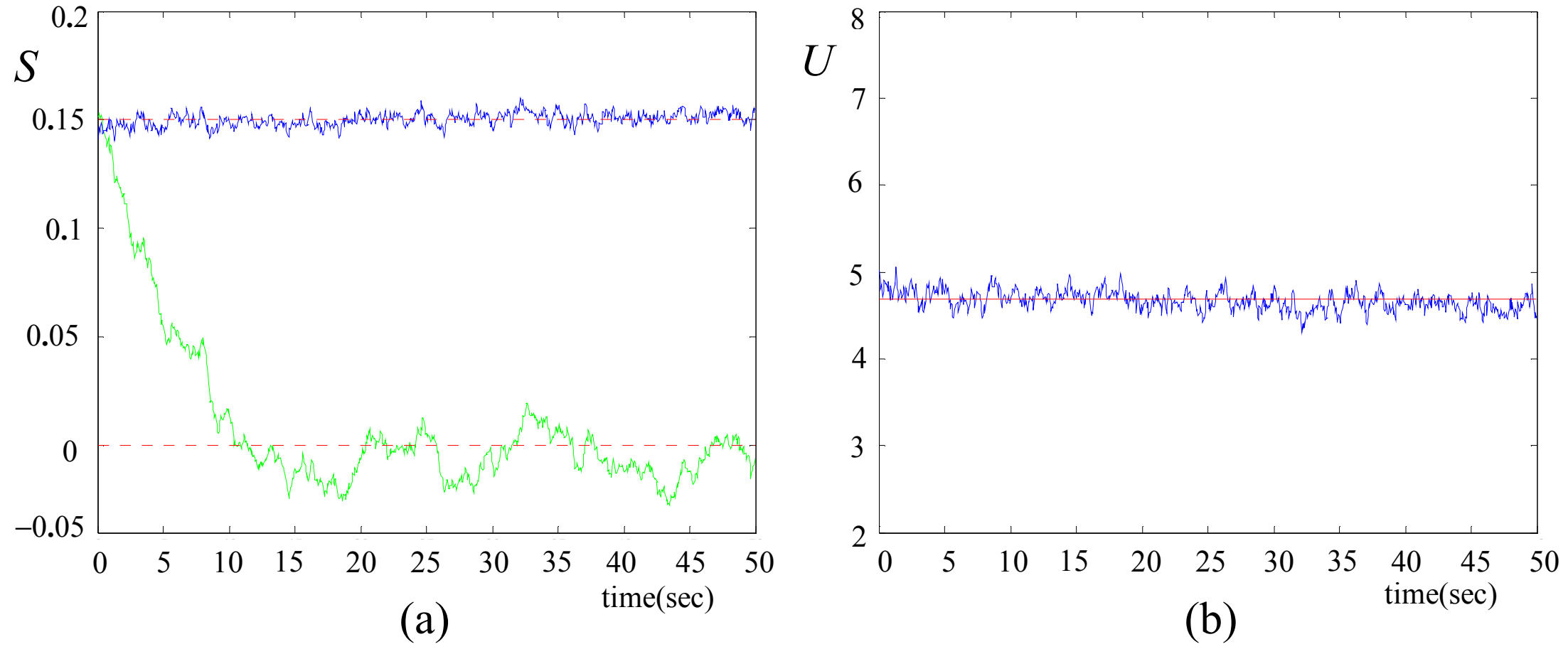
S
0.2
0.15
0.1
0.05
0
−0.05
0 5 10 15 20 25 30 35 40 45 50
time(sec)
(a)
U
8
7
6
5
4
3
2
0 5 10 15 20 25 30 35 40 45 50
time(sec)
(b)

Figure 6.

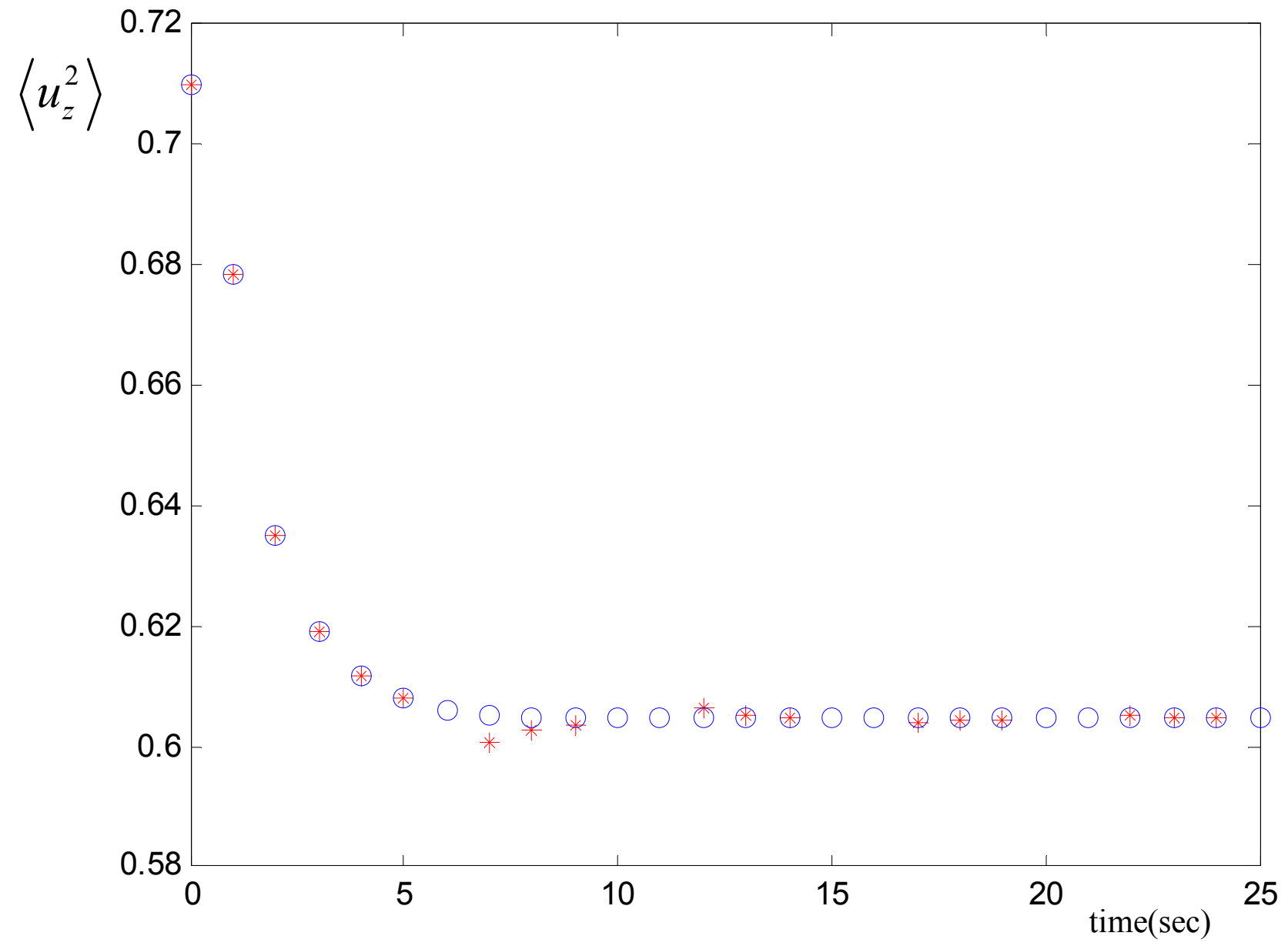

Figure 7.